\documentclass[11pt]{article}
\usepackage{DGfest,epsfig}
\usepackage{times}
\renewcommand\({\left(}
\renewcommand\){\right)}
\renewcommand\[{\left[}
\renewcommand\]{\right]}

\newcommand{\ra}{\rightarrow}

\def\lsim{\raise 0.4ex\hbox{$<$}\kern -0.8em\lower 0.62
ex\hbox{$\sim$}}

\def\gsim{\raise 0.4ex\hbox{$>$}\kern -0.7em\lower 0.62
ex\hbox{$\sim$}}

\def\lbar{{\hbox{$\lambda$}\kern -0.7em\raise 0.6ex
\hbox{$-$}}}

\newcommand\eq[1]{eq.~(\ref{#1})}

\newcommand\p{\partial}

\newcommand\ee{\end{equation}}
\newcommand\be{\begin{equation}}
\def\bea{\begin{array}}
\def\eea{\end{array}}\def\ea{\end{array}}
\newcommand\ees{\end{eqnarray}}
\newcommand\bees{\begin{eqnarray}}
\newcommand\sub[1]{_{\rm #1}}

\def\p1{{\bf p}_1}
\def\p2{{\bf p}_2}
\def\k1{{\bf k}_1}
\def\k2{{\bf k}_2}

\newcommand{\dddM}{\kern 0.2em \raise 1.9ex\hbox{$...$}\kern -1.0em \hbox{$M$}}
\newcommand{\dddQ}{\kern 0.2em \raise 1.9ex\hbox{$...$}\kern -1.0em \hbox{$Q$}}
\newcommand{\dddI}{\kern 0.2em \raise 1.9ex\hbox{$...$}\kern -1.0em\hbox{$I$}}
\newcommand{\dddO}{\kern 0.2em \raise 1.9ex\hbox{$...$}\kern -1.0em
\hbox{${\cal O}$}}
\def\dddz{\raise 1.5ex\hbox{$...$}\kern -0.8em \hbox{$z$}}

\newcommand{\fobs}{f_{\rm obs}}

\newcommand{\DE}{\D E_{\rm rad}}

\newcommand{\msun}{M_{\odot}}

\newcommand{\tret}{t_{\rm ret}}

\def\o{\omega}

\def\D{\Delta}
\def\p{\partial}

\def\nn{\nonumber}

\def\g{\gamma}

\def\L{\Lambda}

\def\dslash{\hspace{-1mm}\not{\hbox{\kern-2pt $\partial$}}}
\def\Dslash{\not{\hbox{\kern-4pt $D$}}}
\def\pslash{\not{\hbox{\kern-2.1pt $p$}}}
\def\kslash{\not{\hbox{\kern-2.3pt $k$}}}
\def\qslash{\not{\hbox{\kern-2.3pt $q$}}}

\newcommand{\mpl}{M_{\rm Pl}}

\newcommand{\hogw}{h_0^2\Omega_{\rm gw}}

\def\ra{\rightarrow}

\def\be{\begin{equation}}
\def\ee{\end{equation}}
\def\bea{\begin{eqnarray}}
\def\eea{\end{eqnarray}}

\begin{document}
\baselineskip 11.5pt
\title{Gravitational waves and fundamental physics}

\author{ Michele Maggiore }

\address{D\'epartement de Physique Th\'eorique,
Universit\'e de Gen\`eve\\
24, quai Ansermet, Gen\`eve, Switzerland }

%

\maketitle\abstracts{
I give an overview of the motivations for gravitational-wave research,
concentrating on the aspects related to ``fundamental'' physics.
}

\section{Introduction}

I am particularly glad to contribute to this Volume in honour of the 70th
birthday of Adriano Di~Giacomo. Adriano has been my teacher, and I learned
much from him. Exactly twenty years ago, I was working for my Diploma thesis
under his supervision, and then I did my PhD with him. The atmosphere of
enthusiasm and energy that we had in his group will certainly stay among my
fond memories.

With Adriano we  worked on  problems that belong to ``fundamental''
physics. In particular, at the center of our interests
was the problem of quark confinement, and more generally
the non-perturbative aspects of QCD.
Presently, I am mostly working on gravitational-wave (GW) physics. 
Then, I   decided to use the opportunity of 
my contribution to this Volume to collect some ideas about
what can we hope to learn about fundamental physics from the observation
and the study of GWs, with the forthcoming generation of GW experiments.

Of course, astrophysics can  certainly be considered fundamental in its own
right, and the direct
detection of GWs emitted by fascinating objects such as neutron
stars or black holes would be a fundamental discovery in itself. Here however
I will discuss ``fundamental'' physics, in the sense that is usually 
given to this word in high-energy physics, i.e. something which has to do with
the basic laws that govern the interaction of matter.

Rather than performing a systematic discussion of all the 
situations where GWs can carry informations on fundamental physics, 
I will concentrate on three  examples that I
consider expecially interesting.

\section{Gravitational waves and quantum chromodynamics}

What could possibly GWs have to do with QCD? The connection
is provided by neutron stars. Neutron stars are remarkable objects from
many points of view. They are the final state of stars which, after the
exhaustion of their nuclear fuel and the subsequent explosion and ejection of
the external layers, remain with a core more massive than the Chandrasekhar
mass $M_{\rm Ch}\simeq 1.4\msun$
(but still the core must be lighter than a critical 
value $M_{\rm bh}<O(2-3)\msun$,
beyond which a black hole instead forms). 
When $M_{\rm Ch}<M_{\rm core} < M_{\rm bh}$,
the core of the star collapses under its own weight, until it reaches a radius
$R\simeq 10$~km, where the self-gravity of the star is now balanced by the
neutron degeneracy pressure. As a consequence, the nuclear matter inside the
star is compressed to extreme densities.  
The internal structure of neutron stars  
depends strongly on the equation of state; however
in the inner core, say $R<O(1)$~km, the density reaches values of order
$1~{\rm GeV}/{\rm fm}^3$.  We are therefore in a regime  governed
by  QCD at high density. This is a non-perturbative  regime, 
which can be used to ask important questions about QCD. 

In particular, what is the true ground state of QCD at low
temperatures? In general, at low temperatures we expect
that  quarks and gluons are confined into hadrons, mainly neutrons and
protons, which in turns are bound together in nuclei. Following this line of
reasonings, one concludes that the true ground state is given by the nucleus
with the largest binding energy per nucleon, which is $^{56}$Fe. However, 
the abundance of  $^{56}$Fe in the Universe is very low. This is explained by
the fact that, when the temperature of the Universe dropped below the
deconfinement temperature, quarks and gluons first were bound together into
neutrons and protons. To synthesize heavier nuclei, it is necessary 
that protons 
overcome their repulsive Coulomb interaction. This energy barrier is larger
for heavier nuclei, so in this
primordial nucleosynthesis only the lightest elements
could be synthesized. Elements heavier than $^7$Li can only be created in
stellar nucleosynthesis, where the huge pressures in the stellar cores forces
the light nuclei to get sufficiently close, overcoming their Coulomb barrier.
Even so, only the heaviest stars can burn their nuclear fuel up to
$^{56}$Fe. This shows that the true ground state of QCD does not necessarily 
correspond to the state of matter that we see around us. The true ground state
is the state of minimum energy, but this can be separated from the initial
state of the Universe by an energy barrier that can be overcome only under
exceptional conditions.

The conditions inside the core of a neutron star (NS) 
are however so extreme, with a density
of order $1~{\rm GeV}/{\rm fm}^3$,  that an energy
barrier of order of the QCD scale can  be overcome, 
and the true ground state
can then be revealed. A particularly interesting possibility is
the strange-quark matter hypothesis,\cite{Bod,Wit} which 
states that the true ground state
  of strong interactions is a deconfined mixture 
  of $u,d,s$ quarks, approximately
  in equal proportions.  We typically expect that,
  energetically, 
deconfined two-flavor quark matter lies about O(100) 
MeV per nucleons above the
  energy per nucleon in nuclei, since this is a typical QCD scale; 
on the other hand, in strange-quark matter this
  could be over-compensated by the fact that now we have three different Fermi
  seas among which a given baryon number can be shared (this is completely
  analogous to the fact that in nuclei it is energetically favorable to have
  approximately an equal number of neutrons and protons, despite the fact that
  the neutrons is heavier than the proton). 
Back-of-the envelope estimates suggest that what we gain opening 
up a third Fermi sea is again of the order of 100 MeV per nucleons. So, 
a quantitative assessment of the
strange matter 
hypothesis is difficult, since it comes from a delicate balance between the
precise numerical values of
non-perturbative quantities. If this hypothesis 
is correct, however, in the core of NS weak interactions
would convert about one third of the $u, d$ quarks into $s$ quarks. 
A star with a quark core is called a hybrid star. Actually,
once initiated, it is quite possible that
this process would extend to the whole star, transforming it into a quark
star.~\cite{Gle}

So, the interior of neutron stars 
is a good place to look for fundamental issues of QCD,
and the question is how can we access it. GWs are a unique probe of this
interior structure, for the following reasons.

A neutron star is a very rigid object,  whose vibrations are
characterized in terms of its normal modes. These normal modes can be
excited, for instance, after a supernova explosion, when the newborn NS
settles down. Furthermore, NSs occasionally undergo catastrophic
rearrangements of their internal structure (crustquake and possibly
corequakes).
A very interesting example of this phenomenon is provided by magnetars,
which are neutron stars with huge magnetic
fields,\cite{DT} of order $10^{14}-10^{15}$G, i.e. 100 to 1000
times stronger than in ordinary pulsars. It is believed that
magnetars provide an explanation for the phenomenon of soft gamma
repeaters (SGR),  x-ray sources that occasionally emit
huge bursts of soft $\g$-rays.
The mechanism invoked to explain the burst activity is that
the magnetic field lines
in magnetars drift through the liquid interior of the NS, stressing
the crust from below and generating strong shear strains. For magnetic fields
stronger than about $10^{14}$~G, these stresses are so large that
they cause the breaking of the  1~km thick NS crust, whose
elastic energy is suddenly released  in a large
starquake, which generates a burst of soft gamma rays.
In has been estimated~\cite{FP2,CDM} that these starquakes can radiate
in GWs an energy $\DE\sim 10^{-10}-10^{-9}\msun c^2$.
For NS at typical galactic distances, bursts
of this type could then  
be detectable already with the next generation of GW
detectors. 
These GWs provide a remarkable probe of the NS interior, for two basic
reasons.

\begin{itemize}

\item When an excited normal mode relaxes by GW
  emission, GWs are emitted (by current quadrupole radiation)
at a frequency equal to the frequency 
of the normal mode. In turn, this
normal mode frequency depends, in a calculable way, 
on the equation of state in the NS interior. Therefore the value of the
GW frequency carries important informations on the internal NS structure.

\item Because of the smallness of gravitational cross-sections,
for GWs even an object  such as the core of a NS is
basically transparent. Therefore, GWs generated inside the core, for instance
as a consequence of a corequake, travel
unaffected outside the NS. This is of course very different from
electromagnetic waves, for which the NS interior is totally opaque, and is an
excellent example of the fact that GW astronomy can potentially open up a
completely new window on the Universe, unaccessible to electromagnetic
observations. 

\end{itemize}

\begin{figure}
\hspace*{1.5cm}\includegraphics[width=0.5\textwidth, angle=270]{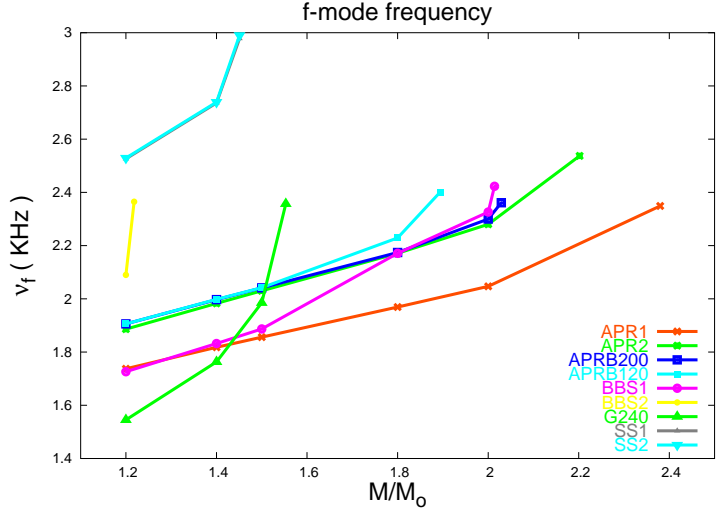}
\begin{flushleft}\caption{The frequency of 
GWs from the $f$-mode for different equations of
  state. The curve labeled 
SS2 corresponds to a quark star, and the (yellow) 
curve BBS2 to a star whose matter
  composition includes the strange baryons $\Sigma^-$ and $\L^0$.
(From Benhar, Ferrari and Gualtieri, Ref.~[7]).
\label{fig:valeria}}
\end{flushleft}
\end{figure}

In Fig.~\ref{fig:valeria} we show the GW frequency emitted by the 
fundamental mode
(or f-mode) of a compact star,
for different equations of state, as a function of the
star mass. It is particularly interesting to see how  GWs emitted by a 
star made entirely of strange quark matter differs from the result for
NS with more
conventional equations of state. The observation of a burst of GWs from a NS  
at a frequency $f$ in the range $2.6-3$~kHz would be a very strong
indication in favor of the existence of strange quark matter.
Estimates of the strength of the GW emission from $f$-modes suggest that
present detectors do not have the sensitivity required for detecting these
waves, unless the source is in our galactic neighborhood; for
a source at a typical galactic distance $r=10$~kpc, these GWs could however be
accessible at advanced detectors.~\cite{BFG}

\section{Compact binaries and the expansion history of the Universe}

A binary system made of two compact stars such as 
neutron stars and/or black holes is a particularly clean system for 
gravitational-wave physics. The stars emit GWs because of their orbital
acceleration. The emission of GWs costs energy, which is drawn from the
orbital energy; then the system slowly spirals inward. As a consequence,   
the orbital
rotational frequency increases, and therefore also the GW frequency increases
with time. This in turns raises the power emitted in GWs so, 
on a timescale of millions
of years, this is a runaway process that ends up in the coalescence of the
system.

\subsection{The Hulse-Taylor binary pulsar}

Experimentally, the decrease of the orbital period of the binary
because of GW emission has been beautifully tested in the famous Hulse-Taylor
binary pulsar (PSR 1913+16). 
This is a system made of two neutron stars,  one of which  is 
observed as a pulsar. Pulsars are extraordinarily stable clocks, which rival in
stability with the best atomic clocks. The arrival time of the pulses on Earth
is however modulated by a number of effects due to 
special and general relativity,
due to the orbital motion of the pulsar around its companion (as well as to
the motion of the detector, on Earth, around the Sun or, more precisely,
around the Solar System Barycenter), 
and to the propagation of the waves in the
gravitational fields of the two NS, and of the Solar System. 
These effects, such as R\"omer, Einstein
and Shapiro time delays, can be accurately computed as a function of the
parameters of the binary system, obtaining the so-called 
timing formula.~\cite{DamourDeruelle} 
Fitting the timing residuals, measured by now over a span of 27 years, to this
timing formula,  the parameters of the orbit of
this binary system have been determined
with great accuracy. To have an idea of the remarkable experimental precision, 
let us mention that the five
so-called Keplerian parameter of the orbit are determined as 
follows~\cite{WT}
\bees
a_p\sin\iota &=&2.3417725(8)\, {\rm s}\, ,\hspace{28mm}
e =0.6171338(4)\, ,\nn\\
T_0&=& 52144.90097844(5) {\rm (MJD)}\, ,\hspace{5mm}
\o_0 =292.54487(8)\, {\rm deg}\, ,\\
&&\hspace*{15mm}P_b = 0.322997448930 (4)\, {\rm days}\, .\nn
\ees
Here $a_p$ is the semimajor axis of the pulsar (in seconds), 
$\iota$ the inclination angle of the orbit with respect to the line of sight,
$e$ the eccentricity of the orbit, $T_0$ a time of passage at periastron 
(Mean Julian Day), $\o_0$ is the periastron angle measured with respect to the
line of nodes, and $P_b$ the orbital period. Furthermore, three so-called
post-Keplerian parameters have been obtained from the timing residuals: 
the rate of advance of the
position of the periastron $\dot{\o}_0$ and  the Einstein parameter $\g$
(which enters in the expression for 
the difference between coordinate time and the pulsar
proper time) are given by
\be
\dot{\o}_0=4.226595(5)\, {\rm deg/yr}\, ,\hspace{15mm}
\g = 0.0042919(8)\, ,
\ee
and the rate of change of the orbital period is
\be
\dot{P}_b =-2.4184(9)\times 10^{-12}\, .
\ee
Assuming the validity of General Relativity, 
from the values of $\dot{\o}_0$ and $\g$ one can obtain the values of the
pulsar mass, $m_p$, and of the companion, $m_c$. The result is
\be
m_p=1.4414(2)\msun\, ,\hspace{5mm}
m_c=1.3867(2)\msun\, .
\ee
Having the masses, and with such a remarkable precision,  
General Relativity
now predicts the value of $\dot{P}_b$ due to the emission of gravitational
radiation. The ratio between the experimental value
$(\dot{P}_{b})_{\rm exp}$ and the value 
$(\dot{P}_{b})_{\rm GR}$ 
predicted by General Relativity turns out to
be  
\be
(\dot{P}_{b})_{\rm exp}/(\dot{P}_{b})_{\rm GR} =1.0013(21)\, .
\ee
This provides a wonderful confirmation of General Relativity, as well as of the
existence of GWs. A confirmation at the level of 30\% comes from another
binary pulsar, PSR~1534+12.\cite{Stairs:2002cw} Recently, it 
has been discovered 
the first double NS
binary in which both neutron stars are observed as pulsars,\cite{Burgay} 
PSR J0737-3039. This system is the most relativistic binary
NS system known, with an orbital period of only 2.4 hr. After just two years of
observation, this system provides confirmation of GW emission at a  level
comparable to  the
Hulse-Taylor binary pulsar. Furthermore, more
post-Keplerian parameters are measurable, providing a confirmation of General
Relativity in strong fields at the 0.1\% level.\cite{Kramer:2005ez}

\subsection{Coalescing binaries as standard candles}

Beside providing the first evidence for GWs,
binary neutron stars are of the
greatest importance for GW research because the last stage of the coalescence
is  among the most interesting sources for GW detectors. In particular, the
frequency of the GW emitted enters into the bandwidth of ground-based
interferometers ($f> O(10)$ Hz) about 17 min before coalescence, and
sweeps up in frequency, until the kHz, where the NS-NS binary coalesce.
A NS-NS merging is a very rare event, on a galactic scale. The expected rate
(determined from the observed population of NS-NS binaries, and dominated by
the recently discovered double pulsar, since this is the system with the
shortest time to merging, 85 Myr), is found to be 
$80^{+210}_{-70}\, {\rm Myr}^{-1}$ per galaxy.~\cite{Lorimer} From this one
finds that
the expected rate for the initial LIGO and VIRGO is 
$35^{+90}_{-30}\times 10^{-3} \,{\rm yr}^{-1}$, while for advanced
interferometers one gets an extremely interesting rate,
$190^{+470}_{-150}\, {\rm yr}^{-1}$, that is, between two
events per day and one event per week!

A remarkable fact about  binary coalescence is that it can provide an
absolute measurement of the distance to the source, something which is
extremely rare, and important, in astronomy. This point can be understood
looking at the waveform of an inspiraling binary;
as long as the system is not at
cosmological distances (so that we can neglect the expansion of the Universe
during the propagation of the wave from the source to the observer) the
waveform of the GW, to lowest order in $v/c$, is
\bees
h_+(t) &=& \frac{4}{r}\(\frac{GM_c}{c^2}\)^{5/3}
\(\frac{\pi f (\tret )}{c}\)^{2/3}\, 
\( \frac{1+\cos^2\iota}{2}\) \cos [\Phi (\tret) ]\, ,\nonumber\\
h_{\times}(t) &=& \frac{4}{r}
\(\frac{GM_c}{c^2}\)^{5/3}
\(\frac{\pi f (\tret )}{c}\)^{2/3}\,
 \cos\iota\, \sin [\Phi (\tret )]\, ,
\label{1times}
\ees
where $h_+$ and $h_{\times}$ are the
amplitudes for the two polarizations of the GW,
$\iota$ is the inclination of the orbit with respect to the line of
sight, 
\be 
M_c =\frac{(m_1m_2)^{3/5}}{(m_1+m_2)^{1/5}}
\ee
is a combination of the masses of the two stars known as the chirp mass,
and 
$r$ is the distance to the source; $f $ is the frequency of the GW, which
evolves in time according to 
\be
\label{1bindf}
\dot{f}= \frac{96}{5}\, \pi^{8/3}\,
\(\frac{GM_c}{c^3}\)^{5/3} f^{11/3}\, ,
\ee
$\tret$ is retarded time, and the phase $\Phi$ is given by 
\be
\Phi (t) = 2\pi \int^t_{t_0} dt' \, f (t')\, .
\ee
For a binary at a  cosmological distance, i.e. 
at redshift $z$,  taking into account the propagation in a
Friedmann-Robertson-Walker Universe, 
these equations are modified in a very simple way:
(1) The frequency that appears in the above formulae is
the frequency measured by the observer, $f_{\rm obs}$, which is red-shifted
with respect to the source frequency $f_s$,
i.e. $f_{\rm obs}=f_s/(1+z)$, and
similarly $t$ and $\tret$ are measured with the observer's clock.
(2) The chirp mass $M_c$ must be replaced by
${\cal M}_c =(1+z) M_c$.
(3) The distance $r$ to the source must be replaced by the luminosity distance 
$d_L(z)$.

Then, the signal received by the observed from
a binary inspiral at redshift $z$, when expressed in terms of the observer
time $t$ , is given by
\bees
h_+(t ) &=& h_c(\tret )\,  \frac{1+\cos^2\iota}{2}\, 
\cos\[ \Phi (\tret )\]
\, ,\nn\\
\label{1hpcosmofin}\\
h_{\times}(t )&=& h_c(\tret )\, \cos\iota\, 
\sin\[ \Phi (\tret )\]\, ,
\ees
where 
\be\label{1hcosmofin}
h_c(t) =\frac{4}{d_L(z)}\(\frac{G{\cal M}_c(z)}{c^2}\)^{5/3}
\(\frac{\pi \fobs(t) }{c}\)^{2/3}\, .
\ee
Let us recall that the luminosity distance $d_L$ of a source
is defined by
\be
{\cal F}=\frac{\cal L}{4\pi d_L^2}\, ,
\ee
where ${\cal F}$ is the  flux
(energy per unit time per unit area)
measured by the observer, and  ${\cal L}$ is 
the absolute luminosity of the source,
i.e. the power that it radiates in its rest frame.
For small redshifts, $d_L$ is related to the present value of the
Hubble parameter $H_0$ and to the deceleration parameter $q_0$  by
\be
\frac{H_0d_L}{c}=z+\frac{1}{2}(1-q_0) z^2+\ldots\, .
\ee
The first term of this expansion give just the Hubble law
$z\simeq (H_0/c) d_L$, which states that
redshifts are proportional to  distances. The term $O(z^2)$ is the
correction to the linear law for moderate redshifts. 
For large redshifts, the
Taylor series is no longer appropriate, and the whole expansion history of
the Universe is encoded in a function $d_L(z)$. As an example, for a 
spatially flat
Universe, one finds
\be
d_L(z)=c \, (1+z)\,\int_0^z\, \frac{dz'}{H(z')}\, ,
\ee
where $H(z)$ is the value of the Hubble parameter at redshift $z$. 
Knowing $d_L(z)$ we can therefore obtain $H(z)$.
This shows that the luminosity distance function
$d_L(z)$ is an extremely important
quantity, which encodes the
whole expansion history of the
Universe.

Now we can understand why coalescing binaries are standard candles.
Suppose that 
we can measure  the amplitudes of both polarizations 
$h_+,h_{\times}$, as well as  $\dot{f}_{\rm obs}$ 
(for  ground-based interferometers, this actually
requires correlations between different detectors). The amplitude of
$h_+$ is $h_c (1+\cos^2\iota)/2$, while the amplitude of
$h_{\times}$ is $h_c\cos\iota$. From their ratio we can therefore obtain the
value of $\cos\iota$, that is, the inclination of the orbit 
with respect to the line of
sight. 
On the other hand, \eq{1bindf} (with the replacement $M_c\ra {\cal M}_c$
mentioned abobe) shows that if we measure 
the value of $\dot{f}_{\rm obs}$ corresponding to a given value of 
$\fobs$, we get ${\cal M}_c$.
Now in the expression for $h_+$ and $h_{\times}$ 
all parameters  have been fixed, except $d_L(z)$.\footnote{It is important
 that the
  ellipticity of the orbit does not enter; it can in fact be shown that, 
by the time that the stars approach the
coalescence stage, angular momentum losses have circularized the orbit to great
accuracy.}
This means that, from the measured
value of $h_+$ (or of $h_{\times}$) we can now read $d_L$. 
If, at the same time, we can measure the redshift $z$ of the source, we have
found a gravitational standard candle, and we can use it to measure the Hubble
constant and, more generally, the evolution of the Universe.\cite{Schutz0}

In a sense, this gravitational standard candle is complementary to the
standard candles that one has  in conventional astrophysics (i.e. from sources
detected electromagnetically). In conventional astrophysics, the
determination of the redshift $z$ of a source is 
straightforward (one simple measures the  redshift of some spectral
line) while the determination of the absolute distance to the source is 
``the''
great problem. Here, on the contrary, there is no theoretical uncertainty 
on the  distance $d_L$, so for the distance
the only error comes from the experimental accuracy
in the measure of the GW. However, the determination of the redshift of the
source can be more difficult. Various possibilities have been proposed. The
simplest, of course, is to see an optical counterpart. In particular a NS-NS
coalescence is also expected to emit a $\g$-ray burst. In this case, 
gravitational
observations give the luminosity distance
$d_L$ of the source while electromagnetic observations could provide 
its redshift $z$. Alternatively, for NS-NS coalescence, we can use the
observational fact that the NS mass spectrum is  strongly peaked around
$1.4\msun$. Then, from the measured value of ${\cal M}_c=(1+z)M_c$, and
assuming $m_1=m_2=1.4\msun$, which gives $M_c\simeq 1.2\msun$, 
we get an estimate of $z$.\cite{Markovic} 
Besides, statistical methods combining  observations of different
coalescences 
have been proposed.\cite{Schutz0}

What sort of ``fundamental'' informations can we get from these measurements? 
At the advanced level, Virgo and LIGO  are expected to detect
at least tens of  NS-NS coalescences per year, up to distances of 
order 2~Gpc, measuring the chirp mass with a precision
that can be better than 0.1\%. The masses of NSs are  typically of
order $1.4\msun$. Stellar-mass black holes, as observed in x-ray binaries,
are in general more massive, typically with masses of order
$10\msun$, and
therefore emit an even more powerful  GW signal during their inspiral
and coalescence.
The coalescence of two black holes, each one with
$10~\msun$, could  be seen by advanced Virgo and advanced LIGO
up to redshifts $z\sim 2-3$.~\cite{Cutler}
Furthermore, the space interferometer LISA, which is expected to fly 
in about 10 years, 
is sensitive to GWs in the mHz region, which corresponds to the
wave emitted by supermassive BH with masses up to $10^6\msun$.  
Nowadays, supermassive BH with masses between $10^6$ and $10^9\msun$ are 
known to exist in the center of most (and probably all) galaxies, including
ours. The coalescence of two supermassive black holes, which could take place
for instance during the collision and merging of two galaxies or of
pre-galactic structure at high redshifts, would be among
the most luminous events in the Universe. Even if the merger rate is poorly
understood,  observations from the Hubble Space Telescope and from x-ray
satellites such as Chandra  have revealed that these merging are not at all
uncommon, out to cosmological distances. LISA could detect them out to
$z\sim 10$,~\cite{Hughes,Vecchio} and is expected 
to measure at least  several events over its mission.

The most important issue that can be addressed with a measure of $d_L(z)$ is
to understand ``dark energy'', the quite mysterious component of the energy
budget of the Universe that manifests itself through an acceleration  of the
expansion of the Universe at high redshift. This has been observed, at 
$z<1.7$,  
using
Type~Ia supernovae as standard candles~\cite{Riess,Perl}.  
A possible concern in
these determinations is the absence of a solid theoretical understanding of the
source. After all, supernovae are complicated phenomena. In particular, one
can be concerned about the possibility of an evolution of the supernovae 
brightness with redshift, and of
interstellar extinction in the host galaxy and in
our Galaxy,
leading to unknown systematics. Gravitational-wave standard candles  could
lead to completely independent determinations, and complement and increase the
confidence in other standard candles,\cite{HH}, as well as extending 
the result to higher
redshifts. In particular, it is of great importance to measure
the equation of state of this dark energy 
component. This can be parametrized in terms 
of the ratio of pressure to density $w=p/\rho$, which can in general 
be a function
of redshift, $w(z)$. The value $w(z)=-1$ corresponds to a cosmological constant
while different values can arise, for instance, from evolving fields.
At present, from Type~Ia supernovae,\cite{Garnavich:1998th}
we have a bound on the average value of $w(z)$ over
values of $z$ up to $z\simeq 1.7$, given by
$\langle w(z)\rangle <-0.55$.
In any case, the answer will have profound implications both in
cosmology and in particle physics.

\section{Stochastic backgrounds and the Big-Bang}

Another possible target of gravitational-wave experiments is given by
stochastic backgrounds of cosmological origin.\cite{MM} 
These are
the gravitational analog of the 2.7~K microwave photon background and 
they can carry unique
informations on the state of the very early Universe and on physics
at correspondingly high energies. To understand this point, it is important to
realize that
a background of relic particles gives a snapshot of the state of
the Universe at a very precise moment, that is,
at the time when these particles decoupled from the
primordial plasma.

In particular, if these particles never reached thermal equilibrium with the 
primordial plasma, they still carry the inprint of the mechanism that created
them. 
In fact, suppose for instance that a stochastic background is generated at
some cosmological epoch, e.g. by a phase transition in the early Universe. If
the particle species in question thermalizes and comes to
equilibrium with the rest
of the primordial plasma, any peculiar feature, for instance of its energy
spectrum, which could have revealed informations 
on the mechanism that generated it, 
will be erased, so for instance the energy spectrum will become  a ``dull''
black-body spectrum. If instead these particles immediately 
decouple from the primordial
plasma, they will arrive to us still carrying all  their information, just
undergoing a redshift because of the expansion of the Universe.

The smaller is the cross section of a particle,
the earlier it decouples. Therefore particles with only gravitational
interactions, like gravitons and possibly other fields predicted by
string theory, decouple much earlier than particles which have  also
electroweak or strong interactions. The condition for
decoupling is that the interaction rate of the process that
maintains equilibrium, $\Gamma$, becomes smaller than the
characteristic time scale, which is given by the Hubble parameter $H$,
\be
\Gamma \ll H \hspace{10mm}\Rightarrow {\rm decoupled}
\ee
(we set  $\hbar =c=1$).
A simple back-of-the-envelope computation shows
that, for gravitons,
\be
\Gamma/ H\sim
(T/\mpl )^3\, ,
\ee
so that gravitons are decoupled
below the Planck scale $\mpl \sim 10^{19}$ GeV, {\it i.e.},
already $10^{-44}$ sec after the big-bang. This means that
a  background of GWs produced in
the very early Universe encodes still today
all the informations about the conditions in which it was created.
In principle, if a sufficiently strong stochastic background has been created
during the Big-Bang, its detection could literally
provide us with a snapshot of the
Big-Bang itself! 

For comparison, the photons
that we observe in the CMB decoupled when the temperature was of
order $T\simeq 0.2$ eV, or $3\cdot 10^5$ yr after the Big Bang.  Therefore,
the photons of the CMB give us a snapshot of the state
of the Universe at $t\sim 3\cdot 10^5$ yr. This
difference in scales  simply reflects the difference in the strength
of the gravitational and electromagnetic interactions.

\subsection{Existing bounds}

The intensity of  a stochastic background of GWs
can be characterized by the dimensionless quantity
\be
\Omega_{\rm gw} (f)=\frac{1}{\rho_c}\,\frac{d\rho_{\rm gw}}{d\log f}\, ,
\ee
where $\rho_{\rm gw}$ is the energy density of the stochastic
background of gravitational waves, $f$ is the frequency, 
$\rho_c=3H_0^2/(8\pi G_N)$ is the present value of the
critical energy density for closing the Universe, and
the present value  of the Hubble parameter  $H_0$ 
is usually written as $H_0=h_0\times 100 $
km/(sec--Mpc), where $h_0=0.71^{+0.04}_{-0.03}$ 
parametrizes the existing experimental
uncertainty.\footnote{Clearly,
it is not very convenient to normalize $\rho_{\rm gw}$
to a quantity, $\rho_c$,
which has an experimental  uncertainty: this uncertainty would appear
in all the subsequent formulas, although it has nothing to do with the 
uncertainties on the GW background itself. 
Therefore, it is customary to
rather characterize the stochastic GW background by
the quantity $\hogw (f)$, which is independent of $h_0$.}

\begin{figure}
\includegraphics[width=0.9\textwidth]{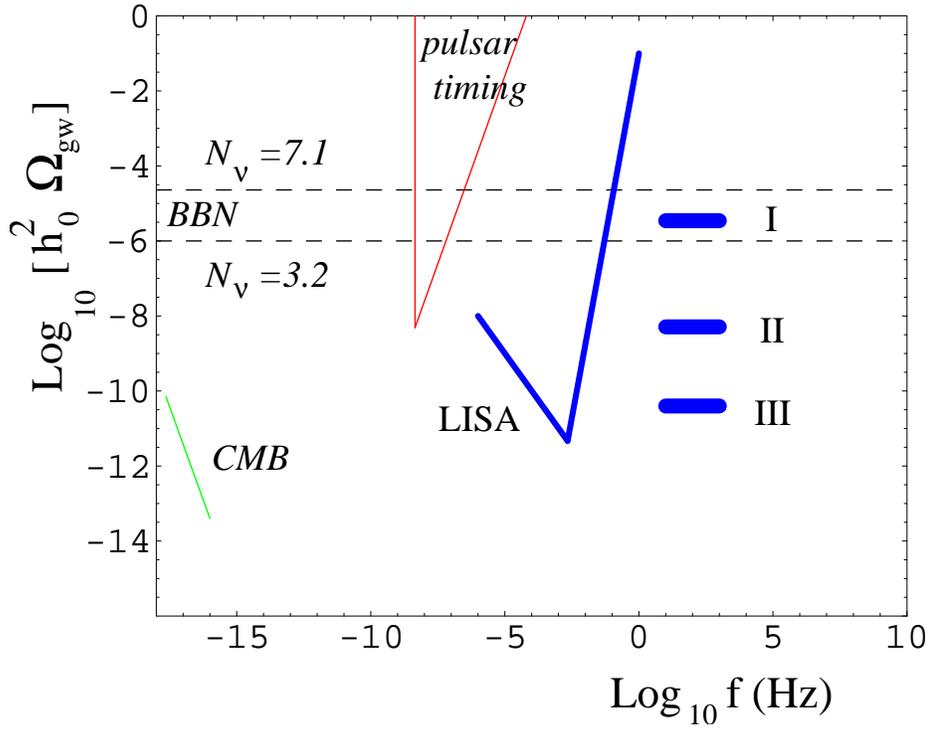}
\begin{flushleft}\caption{Bounds on $\hogw$ from nucleosynthesis
with $N_{\nu}=7.1$ and $N_{\nu}=3.2$ (dashed lines, black), from pulsar timing
(wedge-shaped, red) and from CMB (green), and the
sensitivity  to stochastic GWs
of  LISA, and of the correlation between two 
ground-based interferometers such as VIRGO and LIGO:
(I) two interferometers of first generation.
(II)  two advanced interferometers.
(III) two $3^{rd}$ generation  interferometers.
\label{fig:bounds}}
\end{flushleft}
\end{figure}

At present, we have three major bounds on $\hogw$, illustrated in
Fig.~\ref{fig:bounds}. On the
horizontal axis we plot the GW frequency,  
covering a huge range of frequencies. The lowest value,
$f=10^{-18}$ Hz, corresponds to a wavelength as large as the present Hubble
radius of the Universe; the highest value shown, $f=10^{10}$ Hz, has instead
the following meaning: if we take a graviton produced during the
Planck era, with a typical energy of the order of the Planck or string
energy scale,
and we redshift it to the present time using the standard cosmological
model, we find that today it has a frequency of order 
$10^{10}-10^{12}$ Hz. These values therefore are of order of 
the maximum possible
cutoff of  spectra of GWs produced in the very early Universe. 
The maximum
cutoff for astrophysical processes is of course much lower, of order
10~kHz. So this huge frequency range encompasses all the
GWs that can be considered. 
The three bounds in the figure comes out as follows.

\vspace{3mm}

{\em Nucleosynthesis bound}. The  outcome of big-bang nucleosynthesis  (BBN)
depends on a  balance between
the particle production rates and the expansion rate of the
Universe, measured by the Hubble parameter $H$. Einstein equation
gives $H^2\sim G_N\rho$, where $\rho$ is the total energy density,
including of course $\rho\sub{gw}$. 
Nucleosynthesis successfully predicts the primordial abundances of
deuterium, $^3$He, $^4$He and $^7$Li  
assuming that the only contributions to
$\rho$ come from the particles of the Standard Model, and no GW
contribution. 
Therefore, in order not to spoil the agreement, any further
contribution to $\rho$ at time of nucleosynthesis, including the
contribution of GWs, cannot exceed a maximum
value. The bound is usually written in terms of an effective number of
neutrino species $N_{\nu}$ (i.e. any extra contribution to energy density is
normalized to the energy density 
that would be carried by one neutrino species in
thermal equilibrium). Thus, in the Standard Model, $N_{\nu}=3$, and any
extra form of energy density 
gives further contributions to $N_{\nu}$ (of course, not in general
integer, since the energy of a thermal neutrino is just an arbitrary
normalization scale). In terms of $N_{\nu}$, the bound reads~\cite{MM}
\be\label{nsbound}
\int_{f=0}^{f=\infty}d(\log f)\,\, \hogw (f)\leq
5.6\times 10^{-6}  (N_{\nu}-3)\, .
\ee
The upper limits on $N_{\nu}$ depends on the assumptions made about
nucleosynthesis, and can be as stringent as $N_{\nu}<3.2$. However, if one
invokes a chemical potential for the electron neutrino, this can be relaxed up
to $N_{\nu}<7.1$.~\cite{PDG}

\vspace{3mm}

{\em Bounds from millisecond pulsar}.
Millisecond pulsars are an extremely impressive source of high
precision measurements.\cite{Lorimer} For instance, the  observations of the
first msec pulsar discovered, B1937+21,
give a period of  
$1.557\, 806\, 468\, 819\, 794\, 5(4)$
 ms. As a consequence,
pulsars are also a natural  detector of
GWs, since a GW passing between us and the pulsar 
causes a fluctuation in the time of arrival of the pulse,
proportional to the GW amplitude.
The highest sensitivities can then
be reached for a continuous source, such
as a stochastic background,  after
one or more years of integration, and therefore for 
$f\sim 1/T$ with $T$ equal to a few years, i.e.
$f\sim 10^{-9}-10^{-8}$ Hz. This gives the bound labeled ``pulsar timing''
in Fig.~\ref{fig:bounds}.

\vspace{3mm}

{\em Bounds from CMB}. Another important constraint comes 
from the measurement of the
fluctuation of the temperature of the cosmic microwave background
radiation (CMB). 
The  idea is that
a strong background of GWs at very long wavelengths
produces a stochastic redshift on the frequencies of the photons of
the 2.7K radiation, and therefore a fluctuation 
in their temperature. Indeed, it is quite possible that the observed
anisotropies are partly due to GWs.~\footnote{To discriminate the GW 
contributions from
the effect of scalar perturbations it would be necessary to detect, beside the
already observed $E$-polarization, also the
$B$-polarization of the CMB. This is a target of forthcoming CMB experiments.}
The condition that  a stochastic background 
of GWs does not produce anisotropies in excess of those observed
gives the bound labeled CMB in Fig.~\ref{fig:bounds}.

\subsection{Sensitivity of GW detectors to $\hogw$, 
and theoretical expectations.}

The optimal strategy to detect a stochastic background of GWs is to correlate
the output of two detectors for a time as long as possible. With the present
sensitivities, correlating for four months, and at 90\% confidence level,  the
two LIGO~I detectors would reach $\hogw\simeq 3.5\times 10^{-6}$ (comparable
sensitivities would be reached correlating Virgo with a second nearby
interferometer, such as GEO). At the advanced LIGO level 
one obtains\cite{Ale}
$\hogw\simeq 5.1\times 10^{-9}$ while for so-called third generation
detectors, which are currently under investigation, one could get 
$\hogw\simeq 3.7\times 10^{-11}$ (of course this last figure is more
hypothetical).  LISA, on the other hand, thanks to the fact that it works at
lower frequencies, can reach excellent sensitivities even as a single
detector, with $\hogw\simeq10^{-11}$. These sensitivities 
are shown, together with
the existing bounds, in Fig.~\ref{fig:bounds}. From this figure we see that,
while first generation ground-based
interferometers have marginal chances of detection, at
the level of advanced interferometer and for the space interferometer LISA, we
are penetrating quite deeply into an unexplored region.

The next question, of course, is whether there are theoretical predictions
of stochastic backgrounds of GWs accessible with these
sensitivities. Actually, most predictions of stochastic backgrounds 
of cosmological origin unavoidably face uncertainties due to our ignorance of
early Universe cosmology and/or physics beyond the Standard Model. However,
many examples shows that there are plausible mechanisms for generating
detectable backgrounds\cite{MM}; for instance in string 
cosmology,\cite{GV,Brustein:1995ah}
in a certain range of parameters of the model, is produced 
a GW background that can be observable both
with ground-based interferometers and with LISA,
while at the electroweak  phase transition, in extensions of the Standard
Model, one finds backgrounds that can be detectable at
LISA.\cite{Apreda:2001us} 

The exploration of this new territory might provide us with great rewards.

\end{document}